\documentclass[a4paper,12pt]{article}
\usepackage{epsfig}\parskip 5pt plus 1pt
\usepackage{amsmath}
\usepackage{amssymb}
\usepackage{amsfonts}
\usepackage{fleqn} 
\usepackage{mathrsfs}
\usepackage{cite}
\usepackage{bbm}
\usepackage{graphicx}
\usepackage[footnotesize]{caption}

\newcommand{\CenterEps}[2][1]{\ensuremath{\vcenter{\hbox{\includegraphics[scale=#1]{#2.eps}}}}}
\newcommand{\SuperField}[1]{\hat{#1}}

\def\SU{\text{SU}}

\def\beq{\begin{equation}}
\def\eeq{\end{equation}}
\def\bea{\begin{eqnarray}}
\def\eea{\end{eqnarray}}

\def\<{\left\langle}
\def\>{\right\rangle}
\def\ChargeC{\mathrm{C}} 
\def\chargec{\mathrm{C}}

\allowdisplaybreaks[2]

\begin{document}

\bibliographystyle{OurBibTeX}

\begin{titlepage}

\vspace*{-15mm}
\begin{flushright}
FTUAM 07-07\\ 
IFT-UAM/CSIC 07-18
\end{flushright}
\vspace*{1mm}

\renewcommand{\thefootnote}{\it\alph{footnote}}

\begin{center}
{\bf\Large Flavour-Dependent Type II Leptogenesis} \\[10mm]
{\bf
S.~Antusch\footnote{E-mail: \texttt{antusch@delta.ft.uam.es}}}
\\[5mm]
{\small\it
Departamento de F\'isica Te\'orica C-XI and Instituto
de F\'isica Te\'orica C-XVI,\\
Universidad Aut\'onoma de Madrid, Cantoblanco, E-28049 Madrid, Spain
}
\end{center}
\vspace*{0.40cm}

\begin{abstract}

\noindent
We reanalyse leptogenesis via the out-of-equilibrium decay of 
the lightest right-handed neutrino in type II seesaw scenarios,
taking into account flavour-dependent effects. In the type II 
seesaw mechanism, in addition to the type I seesaw contribution, 
an additional direct mass term for the light neutrinos is present.
We consider type II seesaw scenarios where this additional 
contribution arises from the vacuum expectation value of a Higgs triplet, 
and furthermore an effective model-independent approach.
We investigate bounds on the flavour-specific decay asymmetries, 
on the mass of the lightest right-handed neutrino and on the 
reheat temperature of the early universe, and compare them to the 
corresponding bounds in the type I seesaw framework.  
We show that while flavour-dependent thermal type II leptogenesis 
becomes more efficient for larger mass scale of the light neutrinos, 
and the bounds become relaxed, the type I seesaw scenario for 
leptogenesis becomes more constrained. We also argue that in general,
flavour-dependent effects cannot be ignored when dealing with 
leptogenesis in type II seesaw models.    
\end{abstract}

\end{titlepage}
\newpage
\renewcommand{\thefootnote}{\arabic{footnote}}
\setcounter{footnote}{0}
\section{Introduction}
Leptogenesis \cite{Fukugita:1986hr} is one of the most attractive and minimal 
mechanisms for explaining the observed baryon asymmetry of the universe  
$n_\mathrm{B} /n_\gamma 
\approx (6.0965\,\pm\,0.2055)\,\times\,10^{-10}$ \cite{Spergel:2006hy}. 
A lepton asymmetry is dynamically generated
and then  converted into a baryon asymmetry
due to $(B+L)$-violating sphaleron interactions \cite{Kuzmin:1985mm}
which exist in the Standard Model (SM) and its minimal supersymmetric extension, the MSSM.
Leptogenesis  can be implemented within  the 
type I seesaw scenario \cite{seesaw1}, consisting
of the SM (MSSM)
plus three right-handed Majorana neutrinos (and their superpartners)
with a hierarchical spectrum. 
In thermal leptogenesis \cite{Buchmuller:2005eh}, the lightest of the right-handed  neutrinos is produced by thermal scattering after inflation, and subsequently decays out-of-equilibrium in a lepton number and CP-violating way, thus satisfying Sakharov's constraints \cite{sakharov}.

In models with a left-right symmetric particle content like minimal left-right 
symmetric models, Pati-Salam models or Grand Unified Theories (GUTs) 
based on SO(10), the type I seesaw mechanism is typically   
generalised to a type II seesaw \cite{seesaw2}, 
where an additional direct 
mass term $m_{\mathrm{LL}}^{\mathrm{II}}$ for the light neutrinos is present. 
From a model independent perspective, the type II mass term can 
be considered as an additional contribution to the lowest dimensional 
effective neutrino mass operator. In most explicit models, the type II contribution
stems from seesaw suppressed induced vevs of 
SU(2)$_{\mathrm{L}}$-triplet Higgs fields. 
One motivation for considering the type II seesaw is that it allows to construct unified flavour models for partially degenerate neutrinos in an elegant way, e.g.\ via a type II upgrade \cite{Antusch:2004xd}, which is otherwise difficult to achieve in type I models. 

For leptogenesis in type II seesaw scenarios with SU(2)$_{\mathrm{L}}$-triplet Higgs fields, there are in general two possibilities to generate the baryon asymmetry: via  decays of the lightest right-handed neutrinos or via decays of the  SU(2)$_{\mathrm{L}}$-triplets \cite{O'Donnell:1994am,Ma:1998dx,Hambye:2000ui,Hambye06}. 
In the first case, there are additional one-loop diagrams where 
virtual triplets are running in the loop 
\cite{O'Donnell:1994am,Lazarides:1998iq,Chun:2000dr,Hambye:2003ka,Antusch:2004xy}. 
In the following, we focus on this possibility, and assume hierarchical right-handed neutrino masses (and that the triplets are heavier than $\nu^1_{\mathrm{R}}$).
In this limit, to a good approximation the decay asymmetry depends mainly on the low energy neutrino mass matrix 
$m^{\nu}_{\mathrm{LL}} = m_{\mathrm{LL}}^{\mathrm{I}} +
 m_{\mathrm{LL}}^{\mathrm{II}}$ and on the Yukawa
couplings to the lightest right-handed neutrino and its mass 
\cite{Antusch:2004xy}. It has been shown that type II
leptogenesis imposes constraints on the seesaw parameters, which, in the flavour-independent approximation, differ
substantially from the constraints in the type I case. For instance, 
the bound on the decay asymmetry increases with increasing neutrino mass scale 
\cite{Antusch:2004xy}, in contrast to the type I case where it decreases. 
As a consequence, the lower bound on the mass of the lightest right-handed neutrino from leptogenesis decreases for increasing neutrino 
mass scale \cite{Antusch:2004xy}. 
One interesting application of type II leptogenesis is the possibility to 
improve consistency of classes of unified flavour models with respect to thermal leptogenesis \cite{Antusch:2005tu}.
Finally, since the type II contribution typically does not  
effect washout, there is no bound on the absolute 
neutrino mass scale  
from type II leptogenesis, as has been pointed out in \cite{Hambye:2003ka}.  
For further applications and realisations of type II leptogenesis in specific models of fermion masses and mixings, see e.g.~\cite{TypeIILGModels}.

In recent years, the impact of flavour in thermal leptogenesis has
merited increasing attention~\cite{barbieri} - \cite{res}. 
In fact, the one-flavour
approximation is only rigorously correct when the interactions mediated by 
the charged lepton Yukawa couplings are out of equilibrium. Below a given
temperature (e.g.\ $\mathcal{O}(10^{12}\,\text{GeV})$ in the SM and $(1+\tan^2\beta)\times\mathcal{O}(10^{12}\,\text{GeV})$ in the MSSM), the 
tau Yukawa coupling comes into equilibrium (later followed by the couplings
of the muon and electron). 
Flavour effects are then physical and become manifest, not only at
the level of the generated CP asymmetries, but also regarding the
washout processes that destroy the asymmetries created for each
flavour. In the full computation, 
the asymmetries in each distinguishable flavour are
differently washed out, and appear with distinct weights in the 
final baryon asymmetry.

Flavour-dependent leptogenesis in the type I seesaw scenario has recently been 
addressed in detail by several authors. In particular, 
flavour-dependent effects in leptogenesis have been studied, 
and shown to be relevant, in the two
right-handed neutrino models~\cite{Abada:2006ea} as well as in classes
of neutrino mass models with three right-handed 
neutrinos~\cite{Antusch:2006cw}.
The quantum oscillations/correlations of the asymmetries in 
lepton flavour space have been included in \cite{davidsonetal,Blanchet:2006ch,DeSimone1,DeSimone2} 
and the treatment has 
been generalised to the MSSM \cite{Antusch:2006cw,Antusch:2006gy}. 
Effects of reheating, and constraints on the seesaw parameters from upper bounds on the reheat temperature, have been investigated in \cite{Antusch:2006gy}. 
Leptogenesis bounds on the reheat temperature \cite{Antusch:2006gy} and on the mass of the lightest right-handed neutrino \cite{Antusch:2006gy,Josse-Michaux:2007zj} have also been considered including flavour-dependent effects. 
Strong connections between the low-energy 
CP phases of the $U_\text{MNS}$ matrix and CP violation for flavour-dependent 
leptogenesis have been shown to emerge in certain classes
of neutrino mass models~\cite{Antusch:2006cw} or under the hypothesis
of no CP violation sources associated with the right-handed neutrino 
sector (real $R$) \cite{Blanchet:2006be,Pascoli:2006ie,Branco:2006ce,Pascoli:2006ci}. 
Possible effects regarding the decays of the heavier right-handed neutrinos for leptogenesis have been discussed in this context in \cite{Vives:2005ra,Engelhard:2006yg}, and flavour-dependent effects for resonant leptogenesis were addressed in \cite{res}. 
Regarding the masses of the light neutrinos, assuming hierarchical right-handed neutrinos and considering experimentally allowed light neutrino masses (below about $0.4$ eV), there is no longer a bound on the neutrino mass scale from thermal leptogenesis if flavour-dependent effects are included \cite{Abada:2006ea}.   

In view of the importance of flavour-dependent effects on leptogenesis in the type I seesaw case, it is pertinent to investigate their effects on type II leptogenesis. 
In this paper, we therefore reanalyse leptogenesis via the out-of-equilibrium decay of the lightest right-handed neutrino in type II seesaw scenarios, taking into account flavour-dependent effects. 
We investigate bounds on the decay asymmetries, on the mass of the lightest right-handed neutrino and on the reheat temperature of the early universe, and discuss how increasing the neutrino mass scale affects thermal leptogenesis in the type I and type II seesaw frameworks.

\section{Type I and type II seesaw mechanisms}
\begin{figure}
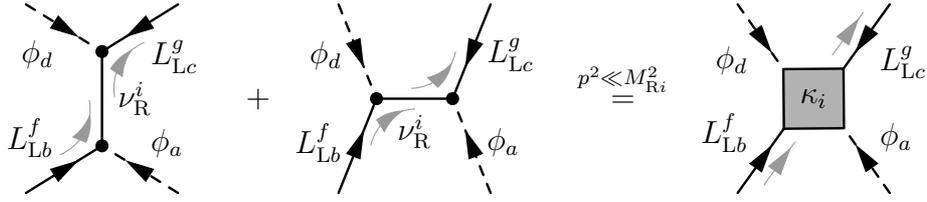

\centering
\CenterEps[1]{typeI}
  \caption{\label{fig:typeI} Generation of the dimension 5 neutrino mass operator in the type I seesaw mechanism.}
\end{figure}

\begin{figure}
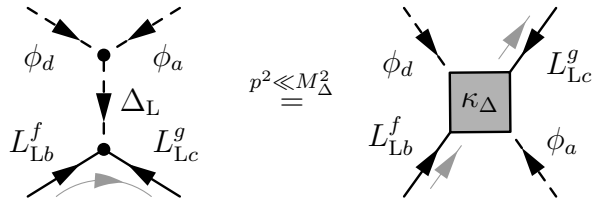
 
\centering
\CenterEps[1]{typeII}
  \caption{\label{fig:typeII}
  Extra diagram generating the dimension 5 neutrino mass operator in the type II seesaw mechanism from a SU(2)$_{\mathrm{L}}$-triplet Higgs field.}
\end{figure}

Motivated by left-right symmetric unified theories, we consider two generic possibilities for explaining the smallness of neutrino masses: via heavy SM (MSSM) singlet fermions (i.e.\ right-handed neutrinos) \cite{seesaw1} and via heavy SU(2)$_{\mathrm{L}}$-triplet Higgs fields \cite{seesaw2}.  
In both cases, the effective dimension five operator for Majorana neutrino masses in the SM or the MSSM, respectively, 
\begin{subequations}\label{eq:The3FormsOfNuMassOp}\begin{eqnarray}
\label{eq:The3FormsOfNuMassOp_SM} \mathscr{L}_{\kappa}^{\mathrm{SM}}  
 &=&\frac{1}{4} 
 \kappa_{gf} \, (\overline{L^\mathrm{C}}^g \cdot \phi)\, 
 \, (L^{f} \cdot \phi) 
  +\text{h.c.} \; , \\
\label{eq:The3FormsOfNuMassOp_MSSM}   \mathscr{L}_{\kappa}^{\mathrm{MSSM}} 
 &=&-\frac{1}{4} 
  {\kappa}^{}_{gf} \, (\SuperField{L}^{g}\cdot
 \SuperField{H}_\mathrm{u}) 
 \, (\SuperField{L}^{f}\cdot \SuperField{H}_\mathrm{u})\, \big|_{\theta\theta} 
 +\text{h.c.} \;,
\end{eqnarray}\end{subequations}
is generated from integrating out the heavy fields. This is illustrated in figures \ref{fig:typeI} and \ref{fig:typeII}. In equation (\ref{eq:The3FormsOfNuMassOp}),  
the dots indicate the $\SU (2)_\mathrm{L}$-invariant product, 
${(\SuperField{L}^{f}\cdot\SuperField{H}_\mathrm{u})} = 
\SuperField{L}_a^{f}(i\tau_2)^{ab} (\SuperField{H}_\mathrm{u})_b$, with $\tau_A$  
$(A\in \{1,2,3\})$ being
the Pauli matrices.   
Superfields are marked by hats.
After electroweak symmetry breaking, the operators of equation (\ref{eq:The3FormsOfNuMassOp}) lead to 
Majorana mass terms for the light neutrinos, 
\begin{equation}
\mathcal{L}_\nu\;=\;-\tfrac{1}{2} m_{LL}^\nu \overline \nu_{\mathrm{L}} 
\nu^{\ChargeC f}_{\mathrm{L}} \;,\;\;
\mbox{with}\;\; m{}_{\mathrm{LL}}^\nu \;=\; - \frac{v_\mathrm{u}^2}{2} (\kappa)^*\;. 
\end{equation}

In the type I seesaw mechanism, it is assumed that only the singlet (right-handed) neutrinos $\nu_{\mathrm{R}i}$ contribute to the neutrino masses. 
With $Y_\nu$ being the neutrino Yukawa matrix in left-right convention,\footnote{The neutrino Yukawa matrix corresponds to $- (Y_\nu)_{f i}({L}^{f}\cdot
 \phi)\,  \nu_\mathrm{\mathrm{R}}^{i} 
$ in the Lagrangian of the SM
and, analogously, to 
$(Y_\nu)_{f i}(\SuperField{L}^{f}\cdot
 \SuperField{H}_\mathrm{u})\,  \SuperField{\nu}^{\chargec i} 
$ in the superpotential of the MSSM (see \cite{Antusch:2004xy} for further details).} 
$M_\mathrm{RR}$ the mass matrix of the right-handed neutrinos and $v_\mathrm{u} = \< \phi^0 \> (= \< H^0_\mathrm{u} \>)$ the vacuum expectation value of the Higgs field which couples to the right-handed neutrinos, the effective mass matrix of the light neutrinos is given by the conventional type I seesaw formula
\begin{eqnarray}
m_{\mathrm{LL}}^{\mathrm{I}} = - v^2_\mathrm{u}\, 
Y_\nu \, M_\mathrm{RR}^{-1}\, Y_\nu^T \; .
\end{eqnarray}

In the type II seesaw mechanism, the contributions to the neutrino mass matrix from both, right-handed neutrinos $\nu_{\mathrm{R}i}$ and Higgs triplet(s) $\Delta_{\mathrm{L}}$, are considered. The additional contribution to the neutrino masses from $\Delta_{\mathrm{L}}$ can be understood in two ways: as another contribution to the effective neutrino mass operator in the low energy effective theory or, equivalently, as a direct mass term after the Higgs triplet obtains an induced small vev after electroweak symmetry breaking (c.f.\  figure \ref{fig:typeII}).   
The neutrino mass matrix in the type II seesaw mechanism has the form 
\begin{eqnarray}
m^\nu_\mathrm{LL} = m_\mathrm{LL}^{\mathrm{II}} + m_\mathrm{LL}^{\mathrm{I}} = m_\mathrm{LL}^{\mathrm{II}}- v_\mathrm{u}^2 Y_\nu M_\mathrm{RR}^{-1} Y_\nu^T \; ,
\end{eqnarray} 
where $m_\mathrm{LL}^{\mathrm{II}}$ is the additional term from the Higgs triplet(s). In left-right symmetric unified theories, the generic size of both seesaw contributions $m^\mathrm{I}_\mathrm{LL}$ and $m^\mathrm{II}_\mathrm{LL}$ is ${\cal O}(v^2_\mathrm{u}/v_\mathrm{B-L})$ where $v_\mathrm{B-L}$ is the B-L breaking scale (i.e.\ the mass scale of the right-handed neutrinos and of the Higgs triplet(s)).

\section{Baryogenesis via flavour-dependent leptogenesis}
Flavour-dependent effects can have a strong impact in baryogenesis via thermal
leptogenesis \cite{barbieri} - \cite{res}. 
The effects are manifest not only in the flavour-dependent CP
asymmetries, but also in the flavour-dependence of scattering processes in the thermal bath, which can destroy a previously produced asymmetry.  

The relevance of the flavour-dependent effects depends on the temperatures at which 
thermal leptogenesis takes place, and thus on 
which interactions mediated by the charged lepton Yukawa
couplings are in thermal equilibrium. 
For example, in the MSSM, for temperatures between circa $(1+\tan^2 \beta)\times 10^{5} \: \mbox{GeV}$ and $(1+\tan^2 \beta)\times 10^{9} \: \mbox{GeV}$, the $\mu$
and $\tau$ Yukawa couplings are in thermal equilibrium and all
flavours in the Boltzmann equations are to be treated separately. 
For $\tan \beta = 30$, this applies for temperatures below about $10^{12}$ GeV and above $10^{8} \: \mbox{GeV}$, a temperature range which is of most interest for thermal leptogenesis in the MSSM. In the SM, in the temperature range between circa $10^{9}$ GeV  and $10^{12}$ GeV, only the $\tau$ Yukawa coupling is in equilibrium and is treated separately in the Boltzmann equations, whereas $\mu$ and $e$ flavours are indistinguishable.  
A discussion of the temperature regimes in the SM and MSSM, where flavour is important, can be found, e.g., in \cite{Antusch:2006cw}.  

We now briefly review the estimation of the produced baryon asymmetry in flavour-dependent leptogenesis.\footnote{For a discussion of approximations which typically enter these estimates, and which also apply to our discussion, 
see e.g.\ section 3.1.3 in \cite{Antusch:2006gy}.} For definiteness, we focus on the temperature range where all flavours are to be treated separately. 
In the following discussion of thermal type II leptogenesis, we will assume that the mass $M_{\Delta_\mathrm{L}}$ of the triplet(s) is much larger than $M_{\mathrm{R}1}$. In this limit, the flavour-dependent efficiencies calculated in the type I seesaw scenario can also be used in the type II framework.
The out-of-equilibrium decays of the heavy right-handed (s)neutrinos $\nu_\mathrm{R}^1$ and $\widetilde \nu_\mathrm{R}^1$  
give rise to flavour-dependent asymmetries in the (s)lepton sector,
which are then partly transformed via sphaleron conversion 
into a baryon asymmetry
$Y_B$.\footnote{In the following, 
$Y$ will always be used for quantities which are normalised to the
entropy density $s$. The quantities normalised with respect to the photon density 
can be obtained using the relation $s/n_\gamma\approx 7.04 k$.}
The final baryon asymmetry can be calculated as
\begin{eqnarray}\label{Eq:YB3f}
Y^\mathrm{SM}_B &=& \frac{12}{37} \sum_f  Y^\mathrm{SM}_{\Delta_f } \; ,\\
Y^\mathrm{MSSM}_B &=& \frac{10}{31} \sum_f  \hat Y^\mathrm{MSSM}_{\Delta_f }\;, 
\end{eqnarray} 
where $\hat Y_{\Delta_f }\equiv Y_B/3 - Y_{L_f }$ are the
total (particle and sparticle) $B/3 - L_f $ asymmetries, with 
$Y_{L_f }$ the lepton number densities in the flavour {\small $f  =
e, \mu,\tau$}.
The asymmetries 
$\hat Y_{\Delta_f }^\mathrm{MSSM}$ and $Y^\mathrm{SM}_{\Delta_f }$, 
which are conserved by sphalerons
and by the other SM (MSSM) interactions,
are then usually calculated by solving a set of coupled Boltzmann equations,
describing the evolution of the
number densities as a function of temperature.

It is convenient to parameterise the produced asymmetries in terms of 
flavour-specific efficiency factors $\eta_{f }$ and decay asymmetries $\varepsilon_{1,f }$ as
\begin{eqnarray}\label{Eq:eta_aa}
Y^\mathrm{SM}_{\Delta_f } &=& \eta^\mathrm{SM}_{f } \,\varepsilon_{1,f } \,
Y^{\mathrm{eq}}_{\nu_\mathrm{R}^1}\;, \\
\label{Eq:eta_aa_MSSM} \hat Y^\mathrm{MSSM}_{\Delta_f } &=& \eta^\mathrm{MSSM}_{f } \,
\left[ 
 \tfrac{1}{2}(\varepsilon_{1,f }+ \varepsilon_{1,\widetilde f }) \, Y^{\mathrm{eq}}_{\nu_\mathrm{R}^1}
 + 
 \tfrac{1}{2}(\varepsilon_{\widetilde 1,f } + \varepsilon_{\widetilde 1,\widetilde f })\, Y^{\mathrm{eq}}_{\widetilde \nu_\mathrm{R}^1}
  \right] .\label{Eq:eta_aa_MSSM}
\end{eqnarray}
  $Y^{\mathrm{eq}}_{\nu_\mathrm{R}^1}$ and $Y^{\mathrm{eq}}_{\widetilde \nu_\mathrm{R}^1}$ are the 
 number densities of the neutrino and sneutrino 
 for $T \gg M_{1}$ if they were in thermal equilibrium, normalised with respect to the entropy density. In the Boltzmann approximation, they are given by
$ Y^{\mathrm{eq}}_{\nu_\mathrm{R}^1} \,\approx \,Y^{\mathrm{eq}}_{\widetilde \nu_\mathrm{R}^1}\,\approx\, 45/( \pi^4 g_* ) $. $g^*$ is the effective number of 
 degrees of freedom, which amounts $106.75$ in the SM and $228.75$ in the MSSM. 

$\varepsilon_{1,f }$, $\varepsilon_{1,\widetilde f }$, $\varepsilon_{\widetilde 1,f }$ and $\varepsilon_{\widetilde 1,\widetilde f }$ are the decay asymmetries for the decay of  
 neutrino into Higgs and lepton,
 neutrino into Higgsino and slepton,
 sneutrino into Higgsino and lepton, and 
 sneutrino into Higgs and slepton, respectively,  
defined by 
\begin{eqnarray}\label{Eq:EpsMSSM_def}
\varepsilon_{1,f } \!\!\!&=&\!\!\!
\frac{
\Gamma_{\nu_\mathrm{R}^1 L_f } - \Gamma_{\nu_\mathrm{R}^1 \overline L_f }
}{
\sum_f  (\Gamma_{\nu_\mathrm{R}^1 L_f } + \Gamma_{\nu_\mathrm{R}^1 \overline L_f })
}\; , \quad
\varepsilon_{1,\widetilde f } =
\frac{
\Gamma_{\nu_\mathrm{R}^1 \widetilde L_f } - \Gamma_{\nu_\mathrm{R}^1
\widetilde{L}_f^*}
}{
\sum_f  (\Gamma_{\nu_\mathrm{R}^1 \widetilde L_f } + \Gamma_{\nu_\mathrm{R}^1
\widetilde{L}_f^*})
}\; , \nonumber  \\
\varepsilon_{\widetilde 1,f } \!\!\!&=&\!\!\!
\frac{
\Gamma_{\widetilde \nu_\mathrm{R}^{*1} L_f } - \Gamma_{\widetilde \nu_\mathrm{R}^1 \overline
  L_f }
}{
\sum_f  (\Gamma_{\widetilde \nu_\mathrm{R}^{*1} L_f } +
\Gamma_{\widetilde \nu_\mathrm{R}^1 \overline L_f })
}\; , \quad
\varepsilon_{\widetilde 1,\widetilde f } =
\frac{
\Gamma_{\widetilde \nu_\mathrm{R}^1 \widetilde L_f } - \Gamma_{\widetilde \nu_\mathrm{R}^{*1}
\widetilde{L}_f^*}
}{
\sum_f  (\Gamma_{\widetilde \nu_\mathrm{R}^1 \widetilde L_f } +
\Gamma_{\widetilde \nu_\mathrm{R}^{*1}
\widetilde{L}_f^*})
}\;.
\end{eqnarray}

The flavour-dependent efficiency factors $\eta_f $ in the SM and in the MSSM are defined by Eqs.~(\ref{Eq:eta_aa}) and (\ref{Eq:eta_aa_MSSM}), respectively. 
As stated above, we assume that the mass $M_{\Delta_\mathrm{L}}$ of the triplet(s) is much larger than $M_{\mathrm{R}1}$. In this limit,  
the efficiencies for flavour-dependent thermal leptogenesis in the type I and type II frameworks are mainly determined by the properties of $\nu_\mathrm{R}^1$, which means in particular that the flavour-dependent efficiencies calculated in the type I seesaw scenario can also be used in the type II framework.
In the definition of the efficiency factor, the equilibrium number densities 
serve as a normalization:   
A thermal population $\nu_{\mathrm{R}1}$ (and $\widetilde \nu_{\mathrm{R}1}$) decaying completely out of equilibrium (without washout effects) would lead to $\eta_f  = 1$.

\begin{figure}[t]
 \centering
 \includegraphics[scale=0.8,angle=0]{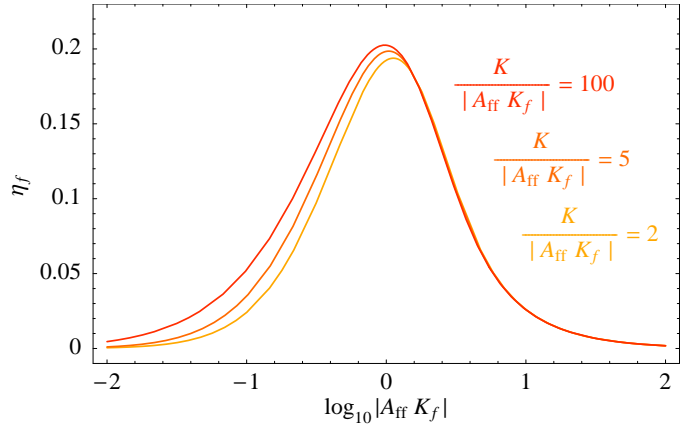}
 \caption{\label{fig:eta}
Flavour-dependent efficiency factor $\eta(A_{f f }K_{f },K)$ in the MSSM as a function of $A_{f f }K_{f }$, for fixed values of $K/|A_{f f }K_{f }| = 2,5$ and $100$, obtained from solving the flavour-dependent Boltzmann equations in the MSSM with zero initial abundance of right-handed (s)neutrinos (figure from \cite{Antusch:2006cw}). $A$ is a matrix which appears in the Boltzmann equations (see \cite{barbieri,Abada:2006ea} for $A$ in the SM and \cite{Antusch:2006cw} for the MSSM case), and which has diagonal elements $|A_{f f }|$ of ${\cal O}(1)$.
The small off-diagonal entries of $A$ have been neglected, which is a good approximation in most cases. In general, however, they have to be included. 
More relevant than the differences in the flavour-dependent efficiency factors for different $K/|A_{f f }K_{f }|$ is that the total baryon asymmetry is the sum of each individual lepton asymmetries, which is weighted by the corresponding efficiency factors.}
\end{figure}

The efficiency factors can be computed by means of the flavour-dependent Boltzmann equations, which can be found for the SM in \cite{barbieri,davidsonetal,nardietal,Abada:2006ea} and for the MSSM in \cite{Antusch:2006cw,Antusch:2006gy}. 
In general, the flavour-dependent efficiencies depend strongly on the washout parameters $\widetilde m_{1,f }$ for each flavour, and on the total washout parameter $\widetilde m_{1}$, which are defined as
\begin{eqnarray}\label{Eq:mtildeaa}\label{eq:mtildeaa}
\widetilde{m}_{1,f  }\, =\,
\frac{v_\mathrm{u}^2 \,\,|(Y_\nu)_{f1}|^2}{M_{\mathrm{R}1}}  \;  , \quad
\widetilde{m}_1\, = \,\sum_f  \widetilde{m}_{1,f  }\; .
\end{eqnarray}
Alternatively, one may use the quantities $K_f ,K$, which are related to $\widetilde{m}_{1,f  },\widetilde{m}_1$ by
\begin{eqnarray}
K_f  \, =\, \frac{\widetilde{m}_{1,f  }}{m^*}\;  , \quad 
K \, =\, \sum_f  K_f  \;,
\end{eqnarray}
with $m^*_\mathrm{SM} \approx 1.08 \times 10^{-3}$ eV  and
$m^*_\mathrm{MSSM} \approx \sin^2(\beta) \times 1.58 \times 10^{-3}$ eV. 
Figure \ref{fig:eta} shows the flavour-specific efficiency factor $\eta_f $ in the MSSM. 
Maximal efficiency for a specific flavour corresponds to $K_f  \approx 1$
 ($\widetilde{m}_{1,f  } \approx m^*$).   

The most relevant difference between the flavour-independent approximation
and the correct flavour-dependent treatment is the fact that in the latter, the total baryon asymmetry is the sum of each individual lepton asymmetries, which is weighted by the corresponding efficiency factor. Therefore, upon summing over the lepton asymmetries, the total baryon number is generically not proportional to the sum over the
CP asymmetries, $\varepsilon_1 = \sum_f  \varepsilon_{1,f }$, 
as in the flavour-independent approximation where the lepton flavour is 
neglected in the Boltzmann equations. 
In other words, in the flavour-independent approximation the total baryon asymmetry is a function of
$\left(\sum_f  \varepsilon_{1,f }\right)\times \eta^\mathrm{ind}\ (\sum_g 
K_{g})$. In the correct flavour treatment
the baryon asymmetry is (approximately) a function of $\sum_f  
\varepsilon_{1,f }\eta\left(A_{f f }K_{f },K\right)$. From this, it is already clear that flavour-dependent effects can have important consequences also in type II leptogenesis. 

The most important quantities for computing the produced baryon asymmetry are thus the decay asymmetries $\varepsilon_{1,f }$ and the efficiency factors $\eta_f $ (which depend mainly on $\widetilde{m}_{1,f  }$ and $\widetilde{m}_1$ (or  $K_{f  }$ and $K$)). 
While the efficiency factors can be computed similarly to the type I seesaw case, important differences between leptogenesis in type I and type II seesaw scenarios arise concerning the decay asymmetries as well as concerning the connection between leptogenesis and seesaw parameters.

\section{Decay asymmetries}

\subsection{Right-handed neutrinos plus triplets}
\begin{figure}
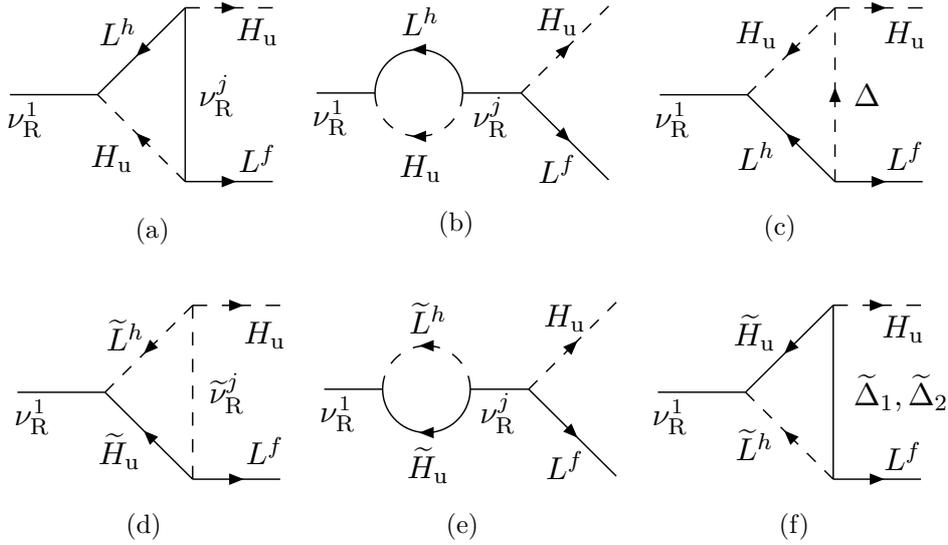
 
\centering
\CenterEps[1]{diagrams}
  \caption{\label{fig:Leptogenesis_Triplet_SUSY}
Loop diagrams in the MSSM which contribute to the decay 
 $\nu^1_{\mathrm{R}}\rightarrow L^f_a H_\mathrm{u}{}_b$ for the case of a 
 type II seesaw mechanism 
 where the direct mass term for the neutrinos stems from the induced vev of a 
 Higgs triplet. 
 In diagram (f), 
 $\widetilde \Delta_1$ and $\widetilde \Delta_2$ are the mass eigenstates
 corresponding to the superpartners of the 
 SU(2)$_{\mathrm{L}}$-triplet scalar fields $\Delta$ and $\bar{\Delta}$.
 The SM diagrams are the ones where no superpartners (marked  
 by a tilde) are involved and where $H_\mathrm{u}{}$ is renamed to the SM Higgs $\phi$.}
\end{figure}

Regarding the decay asymmetry in the  
type II seesaw mechanism, where the direct mass term for the neutrinos 
stems from the induced vev of a Higgs triplet, 
there are new contributions 
from 1-loop diagrams where virtual SU(2)$_{\mathrm{L}}$-triplet scalar fields (or their superpartners) are exchanged in the loop. 
The relevant diagrams for the decay 
 $\nu^1_{\mathrm{R}}\rightarrow L^f_a H_\mathrm{u}{}_b$ 
in the limit $M_1 \ll M_{\mathrm{R}2},M_{\mathrm{R}2},M_\Delta$  
 are shown in figure \ref{fig:Leptogenesis_Triplet_SUSY}. 
Compared to the type I seesaw framework, the new contributions 
are the diagrams (c) and (f).  
The calculation of the corresponding decay asymmetries for each lepton flavour yields 
\begin{subequations}\label{eq:DecayAsTriplet}\begin{eqnarray}\label{eq:EffDecayAss_MinTypeIIScenarios}
 \varepsilon^{(a)}_{1,f} &=& \frac{1}{8 \pi} 
 \frac{\sum_{j\not=1} \mbox{Im}\, 
 [(Y^\dagger)_{1f} (Y^\dagger_\nu Y_\nu)_{1j} (Y^T)_{j f}]}{
 (Y_\nu^\dagger Y_\nu)_{11}} 
  \sqrt{x_j} \left[1 - (1+x_j) \ln \left(\! \frac{x_j +1}{x_j}\!\right) \!\right]
 ,\\
 \varepsilon^{(b)}_{1,f} &=& \frac{1}{8 \pi} 
 \frac{\sum_{j\not=1} \mbox{Im}\, [(Y^\dagger)_{1f} (Y^\dagger_\nu Y_\nu)_{1j} (Y^T)_{j f}]}{
 (Y_\nu^\dagger Y_\nu)_{11}} 
 \, \sqrt{x_j} \,\left[  \frac{1}{1-x_j} \right]
 ,\\
\varepsilon^{(c)}_{1,f} &\!\!=\!\!& -\frac{3}{8 \pi} 
 \frac{M_{\mathrm{R}1}}{v^2_\mathrm{u}}
  \frac{\sum_{g}\mbox{Im}\, [(Y^*_\nu)_{f1} (Y^*_\nu)_{g1} 
 (m_{\mathrm{LL}}^{\mathrm{II}})_{fg}]}{(Y_\nu^\dagger Y_\nu)_{11}}
  \, y \,\left[ - 1 + y \ln \left( \frac{y +1}{y}\right) \!\right]
 ,\\ 
 \varepsilon^{(d)}_{1,f} &=& \frac{1}{8 \pi} 
 \frac{\sum_{j\not=1}\mbox{Im}\, [(Y^\dagger)_{1f} (Y^\dagger_\nu Y_\nu)_{1j} (Y^T)_{j f}]}{
 (Y_\nu^\dagger Y_\nu)_{11}} 
 \, \sqrt{x_j} \,\left[ - 1 + x_j \ln \left( \frac{x_j +1}{x_j}\right) \right]
 ,\\
  \varepsilon^{(e)}_{1,f} &=& \frac{1}{8 \pi} 
 \frac{\sum_{j\not=1} \mbox{Im}\, [(Y^\dagger)_{1f} (Y^\dagger_\nu Y_\nu)_{1j} (Y^T)_{j f}]}{
 (Y_\nu^\dagger Y_\nu)_{11}} 
 \, \sqrt{x_j} \,\left[  \frac{1}{1-x_j} \right], \\
\varepsilon^{(f)}_{1,f} &\!\!=\!\!& -\frac{3}{8 \pi} 
 \frac{M_{\mathrm{R}1}}{v^2_\mathrm{u}}
  \frac{\sum_{g}\mbox{Im}\, [(Y^*_\nu)_{f1} (Y^*_\nu)_{g1} 
 (m_{\mathrm{LL}}^{\mathrm{II}})_{fg}]}{(Y_\nu^\dagger Y_\nu)_{11}}
 \, y \,\left[1 - (1+y) \ln \left( \!\frac{y +1}{y}\!\right) \!\right]
 , 
\end{eqnarray}\end{subequations}
where $y := M^2_\Delta / M^2_{\mathrm{R}1}$ and $x_j := M^2_{\mathrm{R}j} / M^2_{\mathrm{R}1}$ for $j \not= 1$ and where we assume hierarchical right-handed neutrino masses and 
$M_\Delta \gg M_{\mathrm{R}1}$.\footnote{Integrating out the heavy particles $\nu_\mathrm{R}^2,\nu_\mathrm{R}^3,\Delta$ (and their superpartners) in figure \ref{fig:Leptogenesis_Triplet_SUSY} leads to an effective approach involving the dimension 5 neutrino mass operator (c.f.\ figures 1, 2 and \ref{fig:EffLeptogenesis_SUSY}), as will be discussed in section 4.2. 
We note that there are additional diagrams not shown in figure \ref{fig:Leptogenesis_Triplet_SUSY} (since they are generically suppressed for $M_1 \ll M_{\mathrm{R}2},M_{\mathrm{R}2},M_\Delta$) which are related to the dimension 6 operator containing two lepton doublets, two Higgs doublets and a derivative.}

The MSSM results for the type II contributions have been derived in \cite{Antusch:2004xy}. In the SM, the results in \cite{Antusch:2004xy} correct the previous 
result of \cite{Hambye:2003ka} by a factor of $-3/2$. In equation (\ref{eq:DecayAsTriplet}) they have been generalised to the flavour-dependent case. 
The results for the contributions to the decay asymmetries from the triplet in 
the SM and from the triplet superfield in the MSSM are
\begin{subequations}\begin{eqnarray}
\varepsilon^{\mathrm{SM},\mathrm{II}}_{1,f}  &=& 
\varepsilon^{(c)}_{1,f}\; , \\
\varepsilon^{\mathrm{MSSM},\mathrm{II}}_{1,f}  &=& 
\varepsilon^{(c)}_{1,f} + \varepsilon^{(f)}_{1,f}
\; .
\end{eqnarray}\end{subequations}
In the MSSM, we furthermore obtain 
\begin{eqnarray}
\varepsilon{}^{\mathrm{MSSM},\mathrm{II}}_{1,f} = 
 \varepsilon {}\,^{\mathrm{MSSM},\mathrm{II}}_{1,\widetilde f} = 
\varepsilon{}^{\mathrm{MSSM},\mathrm{II}}_{\widetilde 1,f} = 
 \varepsilon {}\,^{\mathrm{MSSM},\mathrm{II}}_{\widetilde 1,\widetilde f}\; .
\end{eqnarray}

The results corresponding to the diagrams (a), (b), (d) and (e) 
which contribute to $\varepsilon_1^\mathrm{I}$ in the type I seesaw in the SM and in the MSSM, have been presented first in \cite{Covi:1996wh}. 
The results for the type I contribution to the decay asymmetries  
in the SM and in the MSSM are
\begin{subequations}\begin{eqnarray}
\varepsilon^{\mathrm{SM},\mathrm{I}}_{1,f}  &=& 
\varepsilon^{(a)}_{1,f} + \varepsilon^{(b)}_{1,f} \; , \\
\varepsilon^{\mathrm{MSSM},\mathrm{I}}_{1,f}  &=& 
\varepsilon^{(a)}_{1,f} + \varepsilon^{(b)}_{1,f} 
+ \varepsilon^{(d)}_{1,f} +
\varepsilon^{(e)}_{1,f}\;.
\end{eqnarray}\end{subequations}
Again, in the MSSM, the remaining decay asymmetries are equal 
to $\varepsilon{}^{\mathrm{MSSM},\mathrm{I}}_{1,f}$:  
\begin{eqnarray}
\varepsilon{}^{\mathrm{MSSM},\mathrm{I}}_{1,f} = 
\varepsilon{}\,^{\mathrm{MSSM},\mathrm{I}}_{1,\widetilde f} = 
\varepsilon{}^{\mathrm{MSSM},\mathrm{I}}_{\widetilde 1,f} = 
\varepsilon{}\,^{\mathrm{MSSM},\mathrm{I}}_{\widetilde 1,\widetilde f}\; .
\end{eqnarray}

Finally, the total decay asymmetries from the decay of $\nu^1_{\mathrm{R}}$ in the type II seesaw, where the direct mass term for the neutrinos stems from the induced vev of a 
Higgs triplet, are given by 
\begin{eqnarray}
\varepsilon^{\mathrm{SM}}_{1,f}  &=&  
\varepsilon^{\mathrm{SM,I}}_{1,f} +  \varepsilon^{\mathrm{SM,II}}_{1,f}\; , \\ 
\varepsilon^{\mathrm{MSSM}}_{1,f} &=&
\varepsilon^{\mathrm{MSSM,I}}_{1,f} +  \varepsilon^{\mathrm{MSSM,II}}_{1,f}\; .
\end{eqnarray}

It is interesting to note that the type I results can be brought to a form which contains the neutrino mass matrix using  
\begin{eqnarray}
 \frac{\sum_{j\not=1} \mbox{Im}\, [(Y^\dagger)_{1f} (Y^\dagger_\nu Y_\nu)_{1j} (Y^T)_{j f}]}{8 \pi\,
 (Y_\nu^\dagger Y_\nu)_{11}} \, \frac{1}{\sqrt{x_j}} =
 -
 \frac{M_{\mathrm{R}1}}{v^2_\mathrm{u}}
  \frac{\sum_{g}\mbox{Im}\, [(Y^*_\nu)_{f1} (Y^*_\nu)_{g1} 
 (m_{\mathrm{LL}}^{\mathrm{I}})_{fg}]}{8 \pi\,(Y_\nu^\dagger Y_\nu)_{11}}\,.
\end{eqnarray}
In the limit $y \gg 1$ and $x_j \gg 1$ for all $j \not= 1$, 
which corresponds to a large gap between the mass  
$M_{\mathrm{R}1}$  and
the masses $M_{\mathrm{R}2}$, $M_{\mathrm{R}3}$ and $M_\Delta$, we obtain the simple results for the flavour-specific decay
asymmetries $\varepsilon^{\mathrm{SM}}_{1,f}$ 
and $\varepsilon^{\mathrm{MSSM}}_{1,f}$ \cite{Antusch:2004xy}
\begin{subequations}\label{Eq:DecayAs_limit}\begin{eqnarray}
\varepsilon^{\mathrm{SM}}_{1,f}  &=& 
  \frac{3}{16 \pi} 
 \frac{M_{\mathrm{R}1}}{v_\mathrm{u}^2} 
 \frac{\sum_{g}\mbox{Im}\, [(Y^*_\nu)_{f1} (Y^*_\nu)_{g1} 
(m^{\mathrm{I}}_{\mathrm{LL}}+ m^{\mathrm{II}}_{\mathrm{LL}})_{fg}]}{
(Y_\nu^\dagger Y_\nu)_{11}}\; ,\\
\varepsilon^{\mathrm{MSSM}}_{1,f} &=&
  \frac{3}{8 \pi} 
 \frac{M_{\mathrm{R}1}}{v_\mathrm{u}^2} 
 \frac{\sum_{g}\mbox{Im}\, [(Y^*_\nu)_{f1} (Y^*_\nu)_{g1} 
(m^{\mathrm{I}}_{\mathrm{LL}}+ m^{\mathrm{II}}_{\mathrm{LL}})_{fg}]}{
(Y_\nu^\dagger Y_\nu)_{11}}\;. 
\end{eqnarray}\end{subequations} 
In the presence of such a mass gap, the calculation  
can also be performed in an effective approach 
after integrating out the two heavy right-handed neutrinos and the heavy 
triplet, as we now discuss.

\subsection{Effective approach to leptogenesis}
Let us now explicitly use the assumption that the lepton asymmetry is generated via the decay of the lightest right-handed neutrino and that all other additional particles, in
particular the ones which generate the type II contribution, are much
heavier than $M_{\mathrm{R}1}$. 
Furthermore, we assume that we can neglect their   
population in the early universe, 
e.g.\ that their masses are much larger than the reheat temperature $T_\mathrm{RH}$ and
that they are not produced non-thermally in a large amount. 
Under these assumptions we can apply an effective approach to leptogenesis, which is independent of the mechanism which generates the additional (type II) contribution to the neutrino mass matrix \cite{Antusch:2004xy}. 

For this minimal effective approach, it is convenient to isolate the type I 
contribution 
from the lightest right-handed neutrino as follows: 
\begin{equation}
(m^{\nu}_{\mathrm{LL}})_{fg} \;=\; - \frac{v_\mathrm{u}^2}{2} \left[ 
 2 (Y_{\nu})_{f1} M_{\mathrm{R}1}^{-1} (Y^T_{\nu})_{1g} + 
 (\kappa'^*)_{fg}\right] .
\end{equation} 
  $\kappa'$ includes type I contributions from the heavier
 right-handed neutrinos, plus any additional (type II) contributions from 
 heavier particles. Examples for realisations of the neutrino mass operator
 can be found, e.g., in \cite{Ma:1998dn}.

At $M_{\mathrm{R}1}$, 
the minimal effective field theory extension of the SM (MSSM) for leptogenesis 
includes the effective neutrino mass operator $\kappa'$ plus 
one right-handed neutrino
$\nu_\mathrm{R}^1$ with mass $M_{\mathrm{R}1}$ and Yukawa couplings 
$(Y_{\nu})_{f1}$ to the lepton doublets $L^f$, defined as 
$
- (Y_\nu)_{f 1}({L}^{f}\cdot
 \phi)\,  \nu_\mathrm{\mathrm{R}}^{1} 
$ in the Lagrangian of the SM
and, analogously, as 
$(Y_\nu)_{f 1}(\SuperField{L}^{f}\cdot
 \SuperField{H}_\mathrm{u})\,  \SuperField{\nu}^{\chargec 1} 
$ in the superpotential of the MSSM.

\begin{figure}
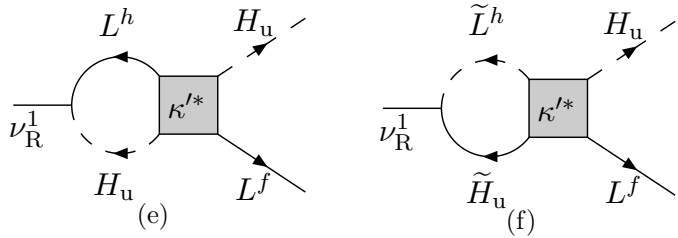

\centering
\CenterEps[1]{effective}
  \caption{\label{fig:EffLeptogenesis_SUSY}
  Loop diagrams contributing to the decay asymmetry via the decay  
 $\nu^1_{\mathrm{R}}\rightarrow L^f_a H_\mathrm{u}{}_b$ in the 
 MSSM with a (lightest) right-handed neutrino 
 $\nu^1_{\mathrm{R}}$ and a  
 neutrino mass matrix determined by $\kappa'$ \cite{Antusch:2004xy}. 
 Further contributions to the generated baryon asymmetry stem from the decay
 of $\nu^1_{\mathrm{R}}$ into slepton and
 Higgsino and from the decays of the sneutrino $\widetilde \nu^1_{\mathrm{R}}$. 
 With $H_\mathrm{u}{}$ 
 renamed to the SM Higgs, the first diagram contributes in the extended SM. }
\end{figure}

The contributions to the decay asymmetries in the effective approach 
stem from the interference of the diagram(s) for the tree-level decay 
of $\nu_{\mathrm{R}1}$ (and $\widetilde \nu_{\mathrm{R}1}$) 
with the loop diagrams containing the effective operator, shown in figure \ref{fig:EffLeptogenesis_SUSY}.
 In the SM, we obtain the simple result \cite{Antusch:2004xy} for the flavour-specific effective decay asymmetries (corresponding to diagram (a) of figure \ref{fig:EffLeptogenesis_SUSY}) 
\begin{eqnarray}\label{eq:EffDecayAss_SM}
\varepsilon{}^{\mathrm{SM}}_{1,f} =
  \frac{3}{16 \pi} 
 \frac{M_{\mathrm{R}1}}{v_\mathrm{u}^2} 
 \frac{\sum_{g}\mbox{Im}\, [(Y^*_\nu)_{f1} (Y^*_\nu)_{g1} 
 (m{}_{\mathrm{LL}}^\nu)_{fg}]}{(Y_\nu^\dagger Y_\nu)_{11}} 
 \;.
 \end{eqnarray}
For the supersymmetric case, diagram (a) and diagram (b) contribute to 
$\varepsilon^{\mathrm{MSSM}}_{1,f}$ and we obtain \cite{Antusch:2004xy}: 
\begin{eqnarray}\label{eq:EffDecayAss_MSSM}
\varepsilon^{\mathrm{MSSM}}_{1,f} =
  \frac{3}{8 \pi} 
 \frac{M_{\mathrm{R}1}}{v_\mathrm{u}^2} 
 \frac{\sum_{g}\mbox{Im}\, [(Y^*_\nu)_{f1} (Y^*_\nu)_{g1} 
 (m{}_{\mathrm{LL}}^\nu)_{fg}]}{(Y_\nu^\dagger Y_\nu)_{11}} \;.
\end{eqnarray}
Explicit calculation furthermore yields 
\begin{eqnarray}\label{eq:SusyDecayAsymmetries}
\varepsilon{}^{\mathrm{MSSM}}_{1,f} = \varepsilon{}\,^{\mathrm{MSSM}}_{1,\widetilde f} =
\varepsilon{}^{\mathrm{MSSM}}_{\widetilde 1,f} =
\varepsilon{}\,^{\mathrm{MSSM}}_{\widetilde 1,\widetilde f}\;.
\end{eqnarray} 
The results are independent of the details of the realisation of the neutrino
mass operator $\kappa'$.  
Note that, since the diagrams where the lightest right-handed neutrino runs in the loop 
do not contribute to leptogenesis, we have written  
$m_{\mathrm{LL}}^\nu = -  {v_\mathrm{u}^2} (\kappa)^*/{2}$ instead of 
$m'{}_{\mathrm{LL}}^\nu:= -  {v_\mathrm{u}^2} (\kappa')^*/{2}$ in the 
formulae in equations (\ref{eq:EffDecayAss_SM}) 
- (\ref{eq:EffDecayAss_MSSM}). The decay asymmetries are directly related to the neutrino mass matrix $m_{\mathrm{LL}}^\nu$.  

For neutrino masses via the type I seesaw mechanism, the results are in agreement 
with the known results \cite{Covi:1996wh}, in the limit 
$M_{\mathrm{R}2},M_{\mathrm{R}3} \gg M_{\mathrm{R}1}$. 
The results obtained in the effective approach are also in agreement with our full theory calculation in the type II scenarios with SU(2)$_{\mathrm{L}}$-triplets in equation 
(\ref{eq:DecayAsTriplet}) \cite{Antusch:2004xy}, in the limit $M_\Delta \gg M_{\mathrm{R}1}$. 

\section{Type II bounds on decay asymmetries and on
$\boldsymbol{M_{\mathrm{R}1}}$}\label{sec:BoundonMR1}
In the limit $M_{\mathrm{R}2},M_{\mathrm{R}3},M_\Delta \gg M_{\mathrm{R}1}$ (or alternatively in the effective approach), upper bounds for the total decay asymmetries in type II leptgenesis, i.e.\ for the sums $|\varepsilon^{\mathrm{SM}}_{1}|= |\sum_f \varepsilon^{\mathrm{SM}}_{1,f}|$ and $|\varepsilon^{\mathrm{MSSM}}_{1}|=|\sum_f \varepsilon^{\mathrm{MSSM}}_{1,f}|$, have been derived in \cite{Antusch:2004xy}. 
For the flavour-specific decay asymmetries $\varepsilon^\mathrm{SM}_{1,f}$ and 
$\varepsilon^\mathrm{MSSM}_{1,f}$, the bounds can readily be obtained as
\begin{eqnarray}\label{eq:BoundsEffDecayAss}\label{eq:BoundEffDecayAss_SM}
|\varepsilon^{\mathrm{SM}}_{1,f}| 
\le
  \frac{3}{16 \pi} 
 \frac{M_{\mathrm{R}1}}{v_\mathrm{u}^2}  m^\nu_\mathrm{max}\; ,
 \quad
\label{eq:BoundEffDecayAss_MSSM} |\varepsilon^{\mathrm{MSSM}}_{1,f}| 
\le
  \frac{3}{8 \pi} 
 \frac{M_{\mathrm{R}1}}{v_\mathrm{u}^2}  m^\nu_\mathrm{max}\;. 
\end{eqnarray}
They are thus identical to the bounds for the total asymmetries. In particular, they 
also increase with increasing mass scale of the light neutrinos. 
Note that, compared to the low energy value, 
the neutrino masses at the scale $M_{\mathrm{R}1}$ are enlarged by
renormalization group running by   
$\approx +20 \% 
$ in the MSSM and $\approx +30 \%
$ in the SM, which raises the bounds on the decay asymmetries by the same values (see e.g.\ figure 4 of \cite{Antusch:2003kp}). 

A situation where an almost maximal baryon asymmetry is generated by thermal leptogenesis can be realised, for example, if the total decay asymmetry nearly saturates its upper bound and if, in addition, the washout parameters $\widetilde m_{1,f}$ for all three flavours approximately take its optimal value. 
Classes of type II seesaw models, where this can be accommodated, have been considered in \cite{Antusch:2004xd,Antusch:2004re,Antusch:2005tu}. In these so-called ``type-II-upgraded'' seesaw models, the type II contribution to the neutrino mass matrix is proportional to the unit matrix (enforced e.g.\ by an SO(3) flavour symmetry or by one of its non-Abelian discrete subgroups). From equation (\ref{Eq:DecayAs_limit}), one can readily see that if the type II contribution ($\propto \mathbbm{1}$) dominates the neutrino mass matrix $m^\nu_{\mathrm{LL}}$, and if $(Y_\nu)_{f1}$ are approximately equal for all flavours $f=1,2,3$ and chosen such that the resulting $\widetilde m_{1,f}$ are approximately equal to $m^*$, we have realised $\eta_f \approx \eta_\mathrm{max}$ for all flavours and simultaneously nearly saturated the bound for the total decay asymmetry.\footnote{We further note that the bound for one of the flavour-specific decay asymmetries can be nearly saturated in this scenario if, for instance, $(Y_\nu)_{21} \approx (Y_\nu)_{31} \approx 0$.}  
     
Assuming a maximal efficiency factor $\eta_\mathrm{max}$ for all flavours in a given scenario, and taking an upper bound for the masses of the light neutrinos $m^\nu_\mathrm{max}$ as well as the observed value $n_\mathrm{B}/n_\gamma \approx (6.0965\,\pm\,0.2055)\,\times\,10^{-10}$ \cite{Spergel:2006hy} for the baryon asymmetry, equation (\ref{eq:BoundsEffDecayAss}) can be transformed into lower type II bounds 
for the mass of the lightest right-handed neutrino \cite{Antusch:2004xy}: 
\label{eq:BoundsMR1}\begin{eqnarray}\label{eq:BoundEffDecayAss_SM}
M^{\mathrm{SM}}_{\mathrm{R}1}
\ge
  \frac{16 \pi}{3} 
 \frac{v_\mathrm{u}^2}{  m^\nu_\mathrm{max}} \frac{n_\mathrm{B}/n_\gamma }{0.99 \cdot
 10^{-2}\,\eta_\mathrm{max}}\; ,
 \quad
\label{eq:BoundEffDecayAss_MSSM} 
M^{\mathrm{MSSM}}_{\mathrm{R}1}
\ge
  \frac{8 \pi}{3} 
 \frac{v_\mathrm{u}^2}{  m^\nu_\mathrm{max}} \frac{n_\mathrm{B}/n_\gamma }{0.92 \cdot
 10^{-2}\,\eta_\mathrm{max}}
 \; .
\end{eqnarray}
The bound on $M_{\mathrm{R}1}$ is lower for a larger neutrino 
mass scale.  

\begin{figure}
 \centering
 \includegraphics[scale=0.8,angle=0]{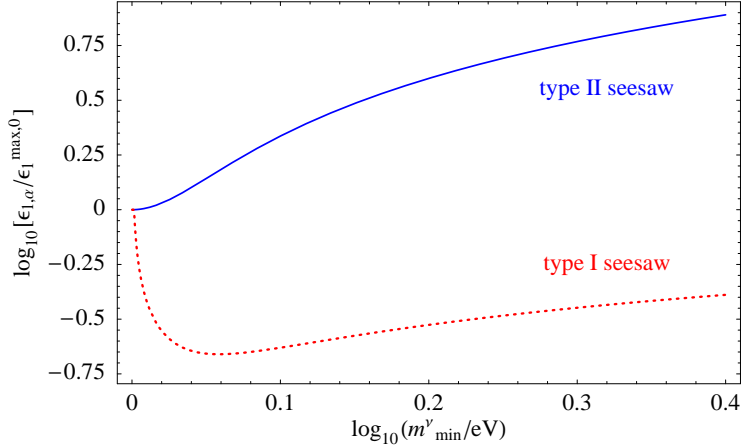}
 \caption{\label{fig:epsbound}
Bound on the decay asymmetry $\varepsilon_{1,f }$ in type II leptogenesis (solid blue line) and type I leptogenesis (dotted red line) 
as a function of the mass of the lightest neutrino $m^\nu_\mathrm{min} := \mbox{min}\,(m_{\nu_1},m_{\nu_3},m_{\nu_3})$ in type I and type II seesaw scenarios 
(see also \cite{Antusch:2006gy}). 
The washout parameter $|A_{f  f }|
\widetilde m_{1,f }$ is fixed to $m^*$ (close to optimal), and the asymmetry is 
normalised to 
$\varepsilon_{1}^{\mathrm{max,0}} = 3\,M_{\mathrm{R}1} \,(\Delta m^2_{31})_{1/2}/(16 \pi\,v_\mathrm{u}^2)$, where $\Delta m^2_{31} \approx 2.5 \times 10^{-3}$ eV$^2$ is the atmospheric neutrino mass squared difference. We have considered the MSSM with $\tan \beta = 30$ as an explicit example. 
}
\end{figure}

The situation in the type II framework differs from the type 
I seesaw case: 
In the latter, the flavour-specific decay asymmetries are
constrained by~\cite{Abada:2006ea} 
\begin{eqnarray}\label{eq:BoundsEffDecayAss_I}\label{eq:BoundEffDecayAss_SM_I}
|\varepsilon^{\mathrm{I,SM}}_{1,f}| 
\le
  \frac{3}{16 \pi} 
 \frac{M_{\mathrm{R}1}}{v_\mathrm{u}^2}  m^\nu_\mathrm{max}\, 
\left(\frac{\widetilde m_{1,f}}{\widetilde m_1}\right)^{\frac{1}{2}} ,
 \quad
\label{eq:BoundEffDecayAss_MSSM} |\varepsilon^{\mathrm{I,MSSM}}_{1,f}| 
\le
  \frac{3}{8 \pi} 
 \frac{M_{\mathrm{R}1}}{v_\mathrm{u}^2}  m^\nu_\mathrm{max}\, 
\left(\frac{\widetilde m_{1,f}}{\widetilde m_1}\right)^{\frac{1}{2}} . 
\end{eqnarray}
Note that compared to the type II bounds, there is an extra factor of 
$\left({\widetilde m_{1,f}}/{\widetilde m_1}\right)^{1/2}$, which depends on the washout parameters. As we shall now discuss, this factor implies that it is not possible to have a maximal decay asymmetry $\varepsilon_{1,f}$ and an optimal washout parameter $\widetilde m_{1,f}$ simultaneously.
Let us recall first that in the type I seesaw, in contrast to the type II case,  
the flavour-independent washout parameter has the lower bound  
\cite{Buchmuller:2003gz}  
\begin{eqnarray}\label{Eq:mtilde_and_m1}
\widetilde m_1 \,\ge\, m^\nu_\mathrm{min} \; ,
\end{eqnarray}
with $m^\nu_\mathrm{min} = \mbox{min}\,(m_{\nu_1},m_{\nu_3},m_{\nu_3})$.  
On the contrary, in the type I and type II seesaw, the flavour-dependent washout parameters $\widetilde m_{1,f}$ are generically not constrained.  
Note that in the flavour-independent approximation,
Eq.~(\ref{Eq:mtilde_and_m1}) leads to a dramatically more restrictive
bound on $\varepsilon_1 = \sum_f  
\varepsilon_{1,f}$~\cite{Davidson:2002qv} for quasi-degenerate light neutrino masses, and finally even to a bound on the neutrino mass scale~\cite{Buchmuller:2003gz}. 
This can be understood from the fact that for $\widetilde
m_{1} \gg m^*$ in the flavour-independent approximation, washout effects 
strongly reduce the efficiency of thermal leptogenesis. Similarly, in the flavour-dependent 
treatment, $\widetilde m_{1,f} \gg m^*$ would lead to a strongly reduced efficiency for this specific flavour. This strong washout for quasi-degenerate light neutrinos can be avoided in flavour-dependent type I leptogenesis, and $\widetilde m_{1,f}\approx m^*$ can realise a nearly optimal scenario regarding washout (c.f.\ figure \ref{fig:eta}). However, we see from equation (\ref{eq:BoundsEffDecayAss_I}) that the decay asymmetries in this case are reduced by a factor of $(m^*/m^\nu_\mathrm{min})^{1/2}$ when compared to the optimal value, leading to a reduced baryon asymmetry. On the other hand, realizing nearly optimal $\varepsilon_{1,f}$ requires $\widetilde m_{1,f} \approx \widetilde m_{1} \ge m^\nu_\mathrm{min}$, leading to large washout effects for quasi-degenerate light neutrinos and even to a more strongly suppressed generation of baryon asymmetry (c.f.\ figure \ref{fig:eta}).
As a consequence, increasing the neutrino mass scale increases the lower bound on $M_{\mathrm{R}1}$ (also in the presence of flavour-dependent effects), in contrast to the type II seesaw case. 

\begin{figure}
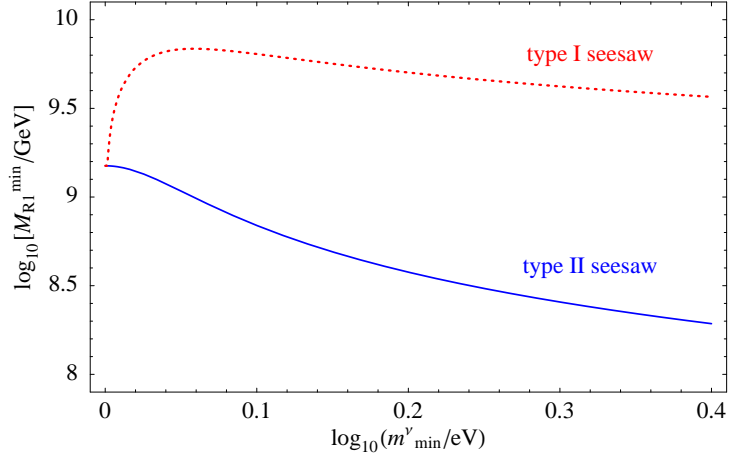

\centering
\CenterEps[0.8]{MR1minTypeIvsTypeII_2}
 \caption{\label{fig:LowerBoundsMR1}
Lower bound on $M_{\mathrm{R}1}$ in type II leptogenesis (solid blue line) and type I leptogenesis (dotted red line) as a function of the mass of the lightest neutrino $m^\nu_\mathrm{min} := \mbox{min}\,(m_{\nu_1},m_{\nu_3},m_{\nu_3})$. 
For definiteness, the MSSM with $\tan \beta = 30$ has been considered as an example.   
 }
\end{figure}

\begin{figure}
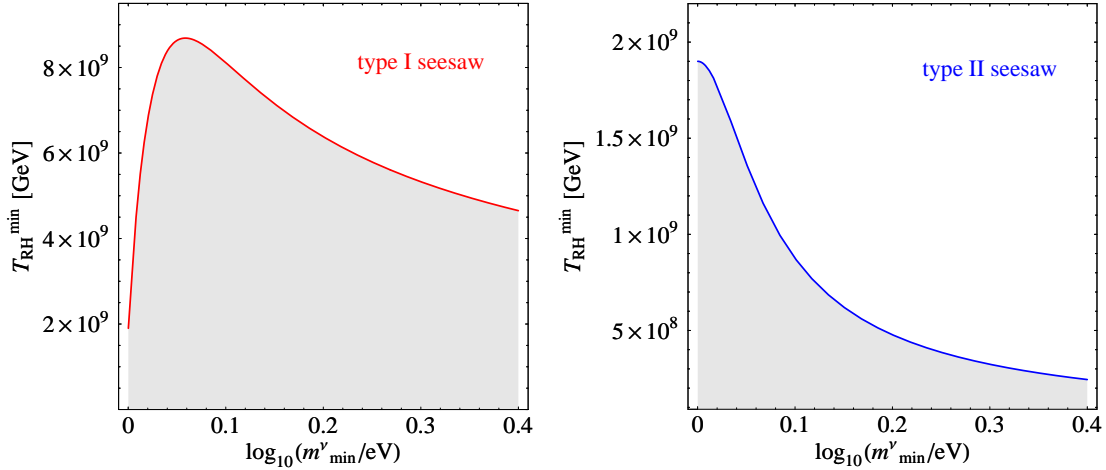

\centering
\CenterEps[0.8]{TRHminTypeI} \;\;\CenterEps[0.8]{TRHminTypeII}
 \caption{\label{fig:LowerBoundsTRH}
Lower bound on the reheat temperature $T_{\mathrm{RH}}$ in type I leptogenesis (left panel) and in type II leptogenesis (right panel)
as a function of the mass of the lightest neutrino $m^\nu_\mathrm{min} = \mbox{min}\,(m_{\nu_1},m_{\nu_3},m_{\nu_3})$, in the MSSM with $\tan \beta = 30$. 
In the grey regions, values of $T_{\mathrm{RH}}$ are incompatible with thermal leptogenesis for the corresponding $m^\nu_\mathrm{min}$.      
 }
\end{figure}

Comparing the type II and type I seesaw cases, in the latter
the baryon asymmetry is suppressed for quasi-degenerate light 
neutrino masses either by a factor $(m^*/m^\nu_\mathrm{min})^{1/2}$ in the decay asymmetries or by a non-optimal washout parameter much larger than $m^*$ (or $K_f \gg 1$,
 c.f.\ figure \ref{fig:eta}). 
The bounds on the decay asymmetries in type I and type II leptogenesis are compared in figure \ref{fig:epsbound}, where $\widetilde m_{1,f}$ has been fixed to $m^*$, close to its optimal value. From figure \ref{fig:epsbound} we see that in the type I case the maximal baryon asymmetry is obtained for hierarchical neutrino masses, whereas in the type II case, increasing the neutrino mass scale increases the produced baryon asymmetry and therefore allows to relax the bound on $M_{\mathrm{R}1}$, as shown in figure \ref{fig:LowerBoundsMR1}.     
In addition, for the same reason, increasing the neutrino mass scale also relaxes the lower bound on the reheat temperature $T_{\mathrm{RH}}$ from the requirement of successful  type II leptogenesis. Including reheating in the flavour-dependent Boltzmann equations as in Ref.~\cite{Antusch:2006gy} (for the flavour-independent case, see \cite{Giudice:2003jh}), we obtain the $m^\nu_\mathrm{min}$-dependent lower bounds on $T_{\mathrm{RH}}$ in type I and type II scenarios shown in figure \ref{fig:LowerBoundsTRH}. While the bound decreases in type II leptogenesis by about an order of magnitude when the neutrino mass scale increases to $0.4$ eV, it increases in the type I seesaw case. 
In the presence of upper bounds on $T_{\mathrm{RH}}$, this can lead to constraints on the neutrino mass scale , i.e.\ on $m^\nu_\mathrm{min} = \mbox{min}\,(m_{\nu_1},m_{\nu_3},m_{\nu_3})$. For instance, with an upper bound $T_{\mathrm{RH}} \le 5 \times 10^9$ GeV, values of $m^\nu_\mathrm{min}$ in the approximate range $\left[0.01\:\mbox{eV}, 0.32\:\mbox{eV}\right]$ would be incompatible with leptogenesis in the type I seesaw framework (c.f.\ figure \ref{fig:LowerBoundsTRH}).

\section{Summary, discussion and conclusions}
We have analysed flavour-dependent leptogenesis via the out-of-equilibrium decay of 
the lightest right-handed neutrino in type II seesaw scenarios, 
where, in addition to the type I seesaw, an additional direct 
mass term for the light neutrinos is present.
We have considered type II seesaw scenarios where this additional 
contribution stems from the vacuum expectation value of a Higgs triplet, and 
furthermore an effective approach, which is independent of the mechanism which generates the additional (type II) contribution to neutrino masses.
We have taken into account flavour-dependent effects, which are 
relevant if thermal leptogenesis takes place at temperatures below circa 
$10^{12} \: \mbox{GeV}$ in the SM and below circa 
$(1+\tan^2 \beta)\times  10^{12} \: \mbox{GeV}$ in the MSSM. 
As in type I leptogenesis, in the flavour-dependent regime the decays of 
the right-handed (s)neutrinos generate asymmetries in in each distinguishable flavour
(proportional to the flavour-specific decay asymmetries $\varepsilon_{1,f}$),
which are differently washed out by scattering processes in the thermal bath, 
and thus appear with distinct weights (efficiency factors $\eta_f $) 
in the final baryon asymmetry.

The most important quantities for computing the produced baryon asymmetry are 
the decay asymmetries $\varepsilon_{1,f }$ and the efficiency factors 
$\eta_f $ (which mainly depend on washout parameters $\widetilde{m}_{1,f  }$ 
and $\widetilde{m}_1 = \sum_f \widetilde{m}_{1,f  }$). 
With respect to the flavour-specific efficiency factors $\eta_f $, in the limit that the mass $M_{\Delta_\mathrm{L}}$ of the triplet is much larger than $M_{\mathrm{R}1}$ (and $M_{\mathrm{R}1} \ll M_{\mathrm{R}2},M_{\mathrm{R}3}$), they can be estimated from the same Boltzmann equations as in the type I seesaw framework.
Regarding the decay asymmetries $\varepsilon_{1,f}$, in the type II seesaw case there are additional contributions where virtual Higgs triplets (and their superpartners) run in the 1-loop diagrams. Here, we have generalised the results of \cite{Antusch:2004xy} to the flavour-dependent case. 
The most important effects of flavour in leptogenesis are a consequence of the fact  that in the flavour-independent 
approximation the total baryon asymmetry is a function of
$\left(\sum_f \varepsilon_{1,f}\right)\times \eta^\mathrm{ind}\ (\sum_g 
\widetilde m_{1,g})$, whereas in the correct flavour-dependent treatment
the baryon asymmetry is (approximately) a function of $\sum_f 
\varepsilon_{1,f} \eta\left(A_{ff} \widetilde m_{1,f}, \widetilde m_1 \right)$.

We have then investigated the bounds on the flavour-specific decay asymmetries $\varepsilon_{1,f}$. In the type I seesaw case, it is known that the bound on 
the flavour-specific asymmetries $\varepsilon^\mathrm{I}_{1,f}$ is substantially relaxed \cite{Abada:2006ea} 
compared to the bound on $\varepsilon^\mathrm{I}_{1} = \sum_f \varepsilon^\mathrm{I}_{1,f}$ \cite{Davidson:2002qv} in the case of a quasi-degenerate spectrum of light neutrinos. For experimentally allowed light neutrino masses below about $0.4$ eV, there is no longer a bound on the neutrino mass scale from the requirement of successful thermal leptogenesis.     
In the type II seesaw case, we have derived the bound on the flavour-specific decay asymmetries $\varepsilon_{1,f} = \varepsilon^\mathrm{I}_{1,f} + \varepsilon^\mathrm{II}_{1,f}$, which turns out to be identical to the bound on the total decay asymmetry $\varepsilon_{1} = \sum_f \varepsilon_{1,f}$. We have compared the bound on the  flavour-specific decay asymmetries in type I and type II scenarios, and found that while the type II bound increases with the neutrino mass scale, the type I bound decreases  (for experimentally allowed light neutrino masses below about $0.4$ eV). The relaxed bound on $\varepsilon_{1,f}$ (figure \ref{fig:epsbound}) leads to a lower bound on the mass of the lightest right-handed neutrino $M_{\mathrm{R}1}$ in the type II seesaw scenario (figure \ref{fig:LowerBoundsMR1}), which decreases when the neutrino mass scale increases. Furthermore, it leads to a relaxed lower bound on the reheat temperature $T_{\mathrm{RH}}$ of the early universe (figure \ref{fig:LowerBoundsTRH}), which helps to improve consistency of thermal leptogenesis with upper bounds on $T_{\mathrm{RH}}$ in some supergravity models. This is in contrast to the type I seesaw scenario, where the lower bound on $T_{\mathrm{RH}}$ from thermal leptogenesis increases with increasing neutrino mass scale. Constraints on $T_{\mathrm{RH}}$ can therefore imply constraints on the mass scale of the light neutrinos also in flavour-dependent type I leptogenesis, although a general bound is absent.

We have furthermore argued that these relaxed bounds on $\varepsilon_{1,f}$ $M_{\mathrm{R}1}$ and $T_{\mathrm{RH}}$ in the type II case can be nearly saturated in an elegant way in classes of so-called ``type-II-upgraded'' seesaw models \cite{Antusch:2004xd}, where the type II contribution to the neutrino mass matrix is proportional to the unit matrix (enforced e.g.\ by an SO(3) flavour symmetry or by one of its non-Abelian subgroups). One interesting application of these type II seesaw scenarios is that the consistency of thermal leptogenesis with unified theories of flavour is improved compared to the type I seesaw case. This effect, investigated in the flavour-independent approximation in \cite{Antusch:2005tu}, is also present analogously in the flavour-dependent treatment of leptogenesis. 
The reason is that if the type II contribution ($\propto \mathbbm{1}$) dominates, the decay asymmetries $\varepsilon_{1,f}$ become approximately equal and the estimate for the produced baryon asymmetry is similar to the flavour-independent case. 
Nevertheless, an accurate analysis of leptogenesis in this scenario requires careful inclusion of the flavour-dependent effects. 
In many applications and realisations of type II leptogenesis in specific models of fermion masses and mixings (see e.g.~\cite{TypeIILGModels}), flavour-dependent effects may substantially change the results and they therefore have to be taken into account.

In summary, type II leptogenesis provides a well-motivated generalisation of the conventional scenario of leptogenesis in the type I seesaw framework. 
We have argued that flavour-dependent effects have to be included in type II leptogenesis, and can change predictions of existing models as well as open up new possibilities for for successful models of leptogenesis. Comparing bounds on $\varepsilon_{1,f}$ $M_{\mathrm{R}1}$ and $T_{\mathrm{RH}}$ in flavour-dependent thermal type I and type II leptogenesis scenarios, we have shown that while type II leptogenesis becomes more efficient for larger mass scale of the light neutrinos, 
and the bounds become relaxed, leptogenesis within the type I seesaw framework becomes more constrained.

\subsection*{Acknowledgments}
I would like to thank Steve~F.~King, Antonio Riotto and Ana~M.~Teixeira for useful discussions and for their collaboration on leptogenesis issues. 
This work was supported by the EU 6$^\text{th}$ Framework Program MRTN-CT-2004-503369 ``The Quest for Unification: Theory Confronts Experiment''.

\end{document}